\documentclass[epj,final]{svjour}
\usepackage{epsfig}

\def\td{t_{\rm d}}
\def\tperp{t_{\perp}}
\def\lan{\left\langle}
\def\ran{\right\rangle}
\def\dx{dx\,}
\def\e{{\rm e}}
\def\virg{\;\;,}
\def\point{\;\,.}
\def\vf{v_{\rm F}}
\def\kf{k_{\rm F}}

\def\ggs{\buildrel\textstyle > \over {\hbox{\raise0.2ex\hbox{$\sim$}}}}
\def\lls{\buildrel\textstyle < \over {\hbox{\raise0.2ex\hbox{$\sim$}}}}
\def\gsim{\,\lower0.75ex\hbox{$\ggs$}\,}
\def\lsim{\,\lower0.75ex\hbox{$\lls$}\,}

\def\Kc   {K_{C}}
\def\Ks   {K_{S}}
\def\Knu  {K_{\nu}}
\def\gncp {G_{\nu + ,{C} +}}
\def\gncm {G_{\nu + ,{C} -}}

\def\gncpm{G_{\nu + ,{C} \pm}}
\def\gnsp {G_{\nu + ,{S} +}}
\def\gnsm {G_{\nu + ,{S} -}}

\def\gnspm{G_{\nu + ,{S} \pm}}
\def\gcs  {G_{{C} p ,{S} p'}}
\def\gcsp {G_{{C} p ,{S} +}}
\def\gcsm {G_{{C} p ,{S} -}}

\def\gcpmsp{G_{{C} \pm ,{S} +}}
\def\gcpmsm{G_{{C} \pm ,{S} -}}
\def\gcpspm{G_{{C} + ,{S} \pm}}
\def\gcmspm{G_{{C} - ,{S} \pm}}
\def\gcps {G_{{C} + ,{S} p}}
\def\gcms {G_{{C} - ,{S} p}}
\def\gcpsp{G_{{C} + ,{S} +}}
\def\gcpsm{G_{{C} + ,{S} -}}

\def\grc  {G_{\rho +, {C} p}}
\def\grcp {G_{\rho +, {C} +}}
\def\grcm {G_{\rho +, {C} -}}
\def\grs  {G_{\rho +, {S} p}}
\def\grsd {G_{\rho +, {S} p'}}
\def\grsp {G_{\rho +, {S} +}}
\def\grsm {G_{\rho +, {S} -}}
\def\gpc  {G_{\sigma +, {C} p}}
\def\gpcp {G_{\sigma +, {C} +}}
\def\gpcm {G_{\sigma +, {C} -}}
\def\gps  {G_{\sigma +, {S} p}}
\def\gpsd {G_{\sigma +, {S} p'}}
\def\gpsp {G_{\sigma +, {S} +}}
\def\gpsm {G_{\sigma +, {S} -}}

\def\ttperp{\widetilde{t}_\perp}

\def\jo #1#2#3#4{#1 {\bf #2}, #4  (#3)}
\def\PR{Phys.\ Rev.}
\def\PRB{Phys.\ Rev.\ B}

\def\SSC{Solid State Commun.}

\def\JPF{J.\ Phys.\ France}

\def\JPSJ{J.\ Phys.\ Soc.\ Jpn.}

\def\PTP{Prog.\ Theor.\ Phys.}

\def\ADV{Adv.\ Phys.}

\def\PHB{Physica B}

\def\IJMPB{Int.\ J.\ Mod.\ Phys.\ B}

\def\SM{Synth.\ Met.}

\def\MCLC{Mol.\ Cryst.\ Liq.\ Cryst.}

\def\EPJB{Eur.\ Phys.\ J.\ B}
\def\SCI{Science}

\begin{document}

\title{Commensurate-incommensurate transition in
two-coupled chains of nearly half-filled electrons}
\titlerunning{Commensurate-incommensurate transition in
two chains of nearly 1/2-filled}

\author{M. Tsuchiizu\inst{1} \and P. Donohue\inst{2}
\and Y. Suzumura\inst{1,3} \and T.Giamarchi\inst{2,4}}

\institute{Department of Physics, Nagoya University, Nagoya
464-8602, Japan
\and
Laboratoire de Physique des Solides, CNRS-UMR
8502, Universit\'e Paris--Sud, B\^at. 510, 91405 Orsay, France
\and
CREST, Japan Science and Technology Corporation (JST), Japan
\and
Laboratoire de Physique Th\'eorique, CNRS UMR 8549, Ecole
Normale Sup\'erieure, 24 Rue Lhomond 75005 Paris, France}

\date{Received:}

\abstract{
We investigate the physical properties of two coupled
chains of electrons, with a nearly half-filled band, as a function
of the interchain hopping $t_\perp$ and the doping. We show that
upon doping, the system undergoes a metal-insulator transition well
described by a commensurate-incommensurate transition. By using
bosonization and renormalization we determine the full phase
diagram of the system, and the physical quantities such as the
charge gap. In the commensurate phase two different regions, for
which the interchain hopping is relevant and irrelevant exist,
leading to a confinement-deconfinement crossover in
this phase. A minimum of the charge gap is observed for values of
$t_\perp$ close to this crossover. At large $t_\perp$ the region
of the commensurate phase is enhanced, compared to a single
chain. At the metal-insulator transition the Luttinger parameter
takes the universal value $K_\rho^*=1$, in agreement with previous
results on special limits of this model.
\PACS{
      {71.10.Hf}{Non-Fermi-liquid ground states, electron
                 phase diagrams and phase transitions in model
                 systems}
      \and
      {71.10.Pm}{Fermions in reduced dimensions
                 (anyons, composite fermions, Luttinger liquid, etc.) }
      \and
      {71.30.+h}{Metal-insulator transitions and other
                 electronic transitions}
      }
}

\maketitle

\section{Introduction}
Organic conductors  Bechgaard salts, which  consist of an array of
one-dimensional chains, show  various ordered states such as
spin-Peierls state,  spin density wave state and superconducting
state at low temperatures \cite{jerome_revue_1d}. Besides these
states of broken symmetry, the normal state above the transition
temperature exhibits a remarkable electronic state associated with
a charge gap. This charge gap is supposed to originate from the
electronic interactions and the commensurate band-filling ($1/4$
filling) of the compound \cite{Giamarchi_physica}.
Indeed in one dimension, commensurate
systems are Mott insulators. Such an
interpretation of the charge gap has received support from recent
optical experiments \cite{dressel_optical_tmtsf,Schwartz} and
transverse conductivity measurements \cite{Jerome_EPJ}. In
addition, since these compounds are quasi-one-dimensional systems
there is a competition between the one dimensional charge gap,
that tends to localize the electrons, and the interchain hopping
tending to make the system three (two) dimensional. This
competition could lead to a confinement-deconfinement transition
responsible for the difference of behavior between the TMTTF and
the TMTSF salts \cite{Giamarchi_physica,Gruner_science}.
Experimentally the  confinement (deconfinement) is found in TMTTF
(TMTSF) salts whose interchain hopping is smaller (larger) than
the charge gap \cite{Gruner_science}.
From a theoretical point of view, the transition should take place when the
interchain hopping renormalized by interactions
is comparable with the single chain gap
\cite{Bourbonnais,Giamarchi_physica,Schwartz}. There is thus still
considerable debate on how such a transition takes place, and what
are the relevant energy scales.

Unfortunately studying an infinite number of coupled chains is
extremely difficult. It is thus worth to investigate these issues
on a finite number of coupled chains, i.e. on ladder systems
\cite{dagotto_ladder_review}. By studying explicitly two-coupled
chains with a half-filled band, it has been shown that the
interchain hopping becomes irrelevant, in the sense of
renormalization group, for a charge gap larger than the
interchain hopping \cite{Suzumura,Tsuchiizu}. It is reasonable to
identify this relevance (irrelevance) of interchain hopping with
the deconfinement (confinement) for an infinite
number of chains since the interchain hopping is renormalized to
zero in the limit of low energy for the irrelevant case. Of
course there are some differences between the simplified ladder
system and the infinite number of chains: (i) for the ladder the
confinement-deconfinement transition is in fact a crossover as far
as the ground state is concerned \cite{LeHur,Donohue}; (ii) in the
ladder both the relevant and irrelevant cases lead to an
insulating state due to a gap in the total charge excitation. Thus
in the ladder to obtain a metallic state, as observed in the
experimental (infinite number of chains) compounds, it would be
necessary to explicitly dope the system and go away from the
commensurate filling. This could be a way to mimic in this
simplified model the small deviation of the commensurate filling
due to warping of the Fermi surface perpendicular to chain
direction \cite{Giamarchi_physica}.

Even if it is known that upon doping the ladder becomes metallic
and the confined region  is suppressed \cite{Suzumura_P}, it is
yet very unclear what are the full properties of the system upon
doping and what are the characteristics of such a metal-insulator
transition. The purpose of the present paper is to examine these
questions in the ladder system. We show that the metal-insulator
transition in the ladder can be accurately described by a
commensurate-incommensurate transition, and we determine its
characteristics. The commensurate-incommensurate transition is
well known in the classical case for
one-dimension \cite{Fukuyama,Okwamoto} or quasi-one-dimensional
\cite{Itakura} system. However the quantum case which corresponds
to interacting electron systems, is known only for one-dimensional
case \cite{Haldane,Schulz1983} where its connections to the Mott
transition have been investigated in details
\cite{Giamarchi_PRB,Mori,Giamarchi_physica}. In the ladder,
although both the half-filled commensurate system (spin ladder)
and the extremely incommensurate case have been widely
investigated \cite{Fabrizio,KR,Nagaosa,Balents,Schulz},
comparatively little has
been done on the metal-insulator transition close to half-filling.
The ladder system with  the umklapp scattering has been examined
in the  chain basis, where the incommensurate phase shows no gap
in the total charge fluctuation \cite{KR}.
Studies using either a mapping on a $SO(8)$ symmetric model
\cite{lin_so8,konik_exact_commensurate_ladder}, onto a hard core
boson system \cite{schulz_mitwochain} or in the large interchain
hopping limit \cite{Ledermann} show a drastic modification of the
universal properties of the metal-insulator transition compared
to the single-chain case.
The  universal properties close to half-filling have been checked
numerically  by DMRG \cite{Siller}.
We use here the
bosonization technique and renormalization group to study the full
problem as a function of the doping and the strength of the
interchain hopping. This allows us to obtain the full phase
diagram and in particular the interplay between the
confinement-deconfinement at commensurate filling and the
metal-insulator transition upon doping the ladder.

The present paper is organized as follows. In Sec.~II, formulation
is given in terms of a bosonized phase Hamiltonian and
renormalization group equations are derived by assuming scaling
invariance for response functions. In Sec.~III, the phase diagram
of the commensurate state and  the incommensurate state  is
calculated by integrating the renormalization group equations. The
charge gap is also estimated and the critical properties of the 
transition are given. In Sec.~IV a summary and a
discussion of the results can be found. Technical details can be
found in the Appendix.

\section{Formulation}

\subsection{System at half-filling}

We consider two-coupled chains of a quarter-filled Hubbard model
 with a  dimerization given by \cite{Tsuchiizu_PHB}
\begin{eqnarray}
{\cal  H} &=& - \sum_{j} \sum_{\sigma = \uparrow, \downarrow}
   \sum_{l=1,2}
   \left[ t + (-1)^{j} \td \right]
   \left(  c_{j, \sigma,l}^{\dagger}
    \,\, c_{j+1, \sigma,l} + \mbox{\rm h.c.} \right)
\nonumber \\ &&{}
-
 2 t_{\perp} \sum_{j} \sum_{\sigma = \uparrow, \downarrow}
\left(
 c_{j, \sigma,1}^{\dagger} \,\,
 c_{j, \sigma,2}  + \mbox{\rm h.c.} \right)
\nonumber \\ &&{} + U \sum_{j,l} \, n_{j, \uparrow,l} \, n_{j,
\downarrow,l} \virg \label{eq:H}
\end{eqnarray}
where $\sigma$ ($=\uparrow, \downarrow$ or $+,-$) and $l$ $(=1,2)$
denote the spin and chain index, respectively, and
$\td$ is the dimerization in the one-dimensional chains.
After the diagonalization of the $\td$-term,
the kinetic term is written as
\begin{eqnarray}
{\cal H}_0 &=& \sum_{k,\sigma,l} \varepsilon_k
  [d_{k,\sigma,l}^\dagger d_{k,\sigma,l}-
     u_{k,\sigma,l}^\dagger u_{k,\sigma,l}]\nonumber \\
  &&-2t_\perp \sum_{k,\sigma} [d_{k,\sigma,1}^\dagger d_{k,\sigma,2}
       +{\rm h.c.}] \\
 && -2t_\perp \sum_{k,\sigma} [u_{k,\sigma,1}^\dagger u_{k,\sigma,2}
       +{\rm h.c.}] \nonumber
\virg
\end{eqnarray}
where $\varepsilon_k =- 2\sqrt{t^2 \cos^2 ka + \td^2 \sin^2 ka}$
\cite{Penc} with  the lattice constant $a$ and $d_{k,\sigma,l}$
($u_{k,\sigma,l}$) is  the fermion operator for the lower (upper)
band on the $l$-th chain. Note that the umklapp scattering is
induced by the dimerization, which has an effect of reducing the
quarter-filled band into an effectively half-filled one. Since we
consider only the lower band, we use half-filling
 instead of quarter-filling in the present paper.
By diagonalizing
 the interchain hopping term with the use of
  $a_{k,\sigma,\zeta}=(-\zeta d_{k,\sigma,1}+d_{k,\sigma,2})/\sqrt{2}$
  ($\zeta=\pm$),
    we obtain ${\cal H}_0= \sum_{k,\sigma,\zeta}
  \varepsilon(k,\zeta)\, a_{k,\sigma, \zeta}^\dagger
   a_{k,\sigma,\zeta}$ where the
  energy dispersion is given by
\begin{eqnarray}
\varepsilon(k,\zeta) &=& - 2\sqrt{t^2 \cos^2 ka + \td^2 \sin^2 ka}
  -2 \tperp \zeta \point
\end{eqnarray}
Here we introduce a linear
  dispersion, $\epsilon(k,\zeta)
  \to \vf (p k- k_{F \zeta})$ where
  $p$ is the index of the branch $p=+(-)$ for
    the right moving (left moving) electrons and
  the new Fermi point is given by
  $k_{{\rm F}\zeta}=\kf -\zeta 2\tperp/\vf$.
We have neglected the $\tperp$-dependence of the
  Fermi velocity.
  After the bosonization of electrons
 around the new Fermi point
 $k_{{\rm F}\zeta}$,
 we introduce the phase variables defined by
  \cite{Tsuchiizu}
\begin{eqnarray}
\theta_{\nu \pm} (x)   \label{eq:phase}
         &=& \frac{1}{\sqrt{2}} \sum_{q\neq 0} \frac{\pi i}{qL}
             \e ^{-\frac{\alpha}{2}|q|-iqx} \sum_{k,\sigma,\zeta}
    f(\sigma,\zeta)
\nonumber \\ &\times&
    \left(  a_{k+q,+,\sigma,\zeta}^\dagger a_{k,+,\sigma,\zeta}
        \pm a_{k+q,-,\sigma,\zeta}^\dagger a_{k,-,\sigma,\zeta}
    \right),
\end{eqnarray}
  where $f(\sigma,\zeta)=1$ (for $\nu= \rho$),
  $\sigma$ (for $\nu= \sigma$), $\zeta$ (for $\nu = C$) and
 $\sigma\zeta$ (for $\nu= S$), respectively.
The phase variables
 $\theta_{\rho +}$ and  $\theta_{\sigma +}$
($\theta_{{\rm C} +}$ and $\theta_{{\rm S}+}$),
     express    fluctuations
 of the total (transverse) charge density and
  spin density. They  satisfy the
  boson  commutation relation given by
$  [\theta_{\nu +}(x),\theta_{\nu' -}(x')]
 = i \pi \delta_{\nu, \nu'}\,{\rm sgn}(x-x') $.
 In terms of these phase variables, the field operator,
 $\psi_{p,\sigma,\zeta}
  =L^{-1/2} \sum_k \e^{ikx} $ $a_{k,p,\sigma,\zeta}$,
 is given by   \cite{Tsuchiizu}
\begin{eqnarray}
\psi_{p,\sigma,\zeta}(x) & = &
 \frac{1}{\sqrt{2\pi \alpha} }
\exp \left( i p k_{{\rm F}\zeta}x
 + i\Theta _{p,\sigma,\zeta} \right)
   \exp \left( i\pi \Xi_{p,\sigma,\zeta} \right)
, \nonumber \\ && \label{eqn:field}
\end{eqnarray}
 where  $\alpha$ is of the order of the lattice constant and
$
\Theta _{p,\sigma,\zeta} = p/(2\sqrt{2})
  [
     \theta_{\rho +} + p\theta_{\rho -}
   + \sigma ( \theta_{\sigma +} + p \theta_{\sigma -} )
   + \zeta    ( \theta_{{\rm C}+} +p \theta_{{\rm C}-} )
   + \sigma \zeta ( \theta_{{\rm S}+} + p\theta_{{\rm S}-} )
                      ]$.
 The phase factor, $\pi \Xi_{p,\sigma,\zeta}$,
  in Eq.~(\ref{eqn:field}),
 is introduced to ensure the anticommutation relation for
   $\psi_{p,\sigma,\zeta}$
  with different $p$, $\sigma$ and $\zeta$.

\subsection{System close to half-filling}

In order to consider the system, which is slightly away from
half-filling, we consider  the following additional term,
\begin{eqnarray}
{\cal H}_\mu &=&
  -\mu \sum_{j,\sigma,l} c_{j,\sigma,l}^\dagger \, c_{j,\sigma,l}
\nonumber \\ & = & - \mu \frac{\sqrt{2}}{\pi} \int dx \,
\partial_x
  \theta_{\rho +} \virg
\label{eq:mu}
\end{eqnarray}
  where $\mu$ is the chemical potential
  and $\mu = 0$ corresponds to the half-filling.
We apply  the transformation $\sqrt{2}\theta_{\rho +} \to
  \sqrt{2}\theta_{\rho+}+q_0 x$
  with  $q_0= 4\mu K_\rho /v_\rho$, which leads to
   a misfit term $q_0 x$ in the cosine term
   expressing  the umklapp scattering.
The phase Hamiltonian is written as,
\begin{eqnarray}
{\cal H} &=& \sum_{\nu = \rho,\sigma,{\rm C},{\rm S}}
\frac{v_\nu}{4\pi} \int \hspace{-1mm} \dx
 \left[
   \frac{1}{K_\nu} \left(\partial \theta_{\nu +} \right)^2
         +  K_\nu  \left(\partial \theta_{\nu -} \right)^2
 \right]
\nonumber \\
&&+
    \frac{g_\rho}{4\pi^2 \alpha^2} \int \hspace{-1mm} \dx
    \biggl[
        \cos \biggl( \sqrt{2}\theta_{{\rm C}+}
                     - \frac{8\tperp}{\vf}x\biggr)
      + \cos \sqrt{2} \theta_{{\rm C}-}
    \biggr]
\nonumber \\ && \hspace{2cm}\times
    \left[
        \cos \sqrt{2} \theta_{{\rm S} +}
      - \cos \sqrt{2} \theta_{{\rm S} -}
    \right] \nonumber \\
&&+
    \frac{g_\sigma}{4\pi^2 \alpha^2} \hspace{-1mm}
     \int \hspace{-1mm} \dx
    \biggl[
        \cos \biggl( \sqrt{2}\theta_{{\rm C} +}
                        -\frac{8\tperp}{\vf}x\biggr)
      - \cos \sqrt{2} \theta_{{\rm C} -}
    \biggr]
\nonumber \\ && \hspace{2cm}\times
    \left[
        \cos \sqrt{2} \theta_{{\rm S} +}
             + \cos \sqrt{2} \theta_{{\rm S} -}
    \right] \nonumber \\
&&+
     \frac{g_{u}}{2\pi^2 \alpha^2} \int \hspace{-1mm}\dx
    \sin \left(\sqrt{2} \theta_{\rho +} + q_0 x \right)
\nonumber \\ && \hspace{1cm}\times
    \biggl[
        \cos \biggl( \sqrt{2}\theta_{{\rm C}+} - \frac{8\tperp}{\vf}x
             \biggr)
      + \cos \sqrt{2} \theta_{{\rm C} -}
\nonumber \\ &&  \hspace{3.cm}
      - \cos \sqrt{2} \theta_{{\rm S} +}
      + \cos \sqrt{2} \theta_{{\rm S} -}
    \biggr]
 \nonumber \\
&&+
    \frac{g_{\perp}}{2\pi^2 \alpha^2}  \int \hspace{-1mm}\dx
    \cos \sqrt{2} \theta_{\sigma +}
\nonumber \\ && \hspace{1cm}\times
    \biggl[
        \cos \biggl( \sqrt{2}\theta_{{\rm C} +}
                     - \frac{8\tperp}{\vf}x\biggr)
      - \cos \sqrt{2} \theta_{{\rm C} -}
\nonumber \\ &&  \hspace{3cm}
      - \cos \sqrt{2} \theta_{{\rm S} +}
      - \cos \sqrt{2} \theta_{{\rm S} -}
    \biggr]
                                   \virg
\label{phase_Hamiltonian}
\end{eqnarray}
where $v_{\rho(\sigma)} = \vf [1+\! (-) \, U/\pi \vf]^{1/2}$,
      $v_{\rm C(S)} = \vf$,
      $K_{\rho(\sigma)} = [1+\!(-) \, U/\pi\vf]^{-1/2}$,
      $K_{\rm C(S)} =1$,
      $g_\rho = -g_\sigma =g_\perp = Ua$ and
  a coupling constant for the umklapp scattering is given by
      $g_3=-Ua(2\td/t)/[1+(\td/t)^2]$ \cite{Penc}.

\subsection{Renormalization group equations}

By  utilizing a renormalization group method,
we analyze Eq.~(\ref{phase_Hamiltonian}) where
the nonlinear terms in Eq.~(\ref{phase_Hamiltonian})
are rewritten as
\begin{equation}
\frac{1}{2\pi^2\alpha^2} \,\, g_{\nu p, \nu' p'}
     \int dx \cos \sqrt{2} \, \overline{\theta}_{\nu p}\,
             \cos \sqrt{2} \, \overline{\theta}_{\nu' p'}  \virg
\end{equation}
  where $\overline{\theta}_{\rho +}=\theta_{\rho+} +q_0 x/\sqrt{2}$,
  $\overline{\theta}_{C +}=\theta_{\rho+} +4 \sqrt{2} \tperp x/\vf$
   and  $\overline{\theta}_{\nu p} = \theta_{\nu p}$ otherwise.
The coupling constants are given by
  $g_{\rho +,C\pm}=\mp g_{\rho+,S\pm}=g_u$,
  $\pm g_{\sigma+,C\pm}=-g_{\sigma+,S\pm}=g_\perp$,
  $\pm g_{C\pm ,S\pm} =(g_\rho+g_\sigma)/2$ and
  $\pm g_{C\pm,S\mp}=-(g_\rho-g_\sigma)/2$,
  where each  coupling constant is   treated
  in the renormalization  group  method.
 The renormalization group equations are derived
   from the  response functions,   which are given by
  $\langle T_\tau \exp[i\theta_{\rho+}(x_1,\tau_1)] \,
                     \exp[-i\theta_{\rho+}(x_2,\tau_2)]\rangle$
  \cite{Giamarchi_JPF},
   By making use of  the scaling of the cutoff
  ($\alpha \to \alpha' =\alpha \e^{dl}$), we obtain
  the  equations  as
 (Appendix A),
\begin{eqnarray}
\frac{d}{dl}
 \ttperp
&=& \ttperp
       - \frac{1}{8}
             \grcp^2
         \Kc  \, F_1 (8\ttperp; q_0 \alpha)
\nonumber \\ &&
       - \frac{1}{8}
         \left( \gpcp^2  + \gcpsp^2 + \gcpsm^2
         \right)
         \Kc   J_1 (8\ttperp ),
\nonumber \\ && \label{eq:dl-tperp}
\\
\frac{d}{dl}
 q_0 \alpha
&=&
  q_0 \alpha
       -  \grcp^2
         K_\rho  \, F_1 (q_0 \alpha;8\ttperp)
\nonumber \\ &&
       - \left( \grcm^2 + \grsp^2 + \grsm^2 \right)
         K_\rho  \, J_1 (q_0 \alpha )
, \nonumber \\ && \label{eq:dl-q0}
\\
\frac{d}{dl}
 K_\rho &=&
- \frac{1}{2} K_\rho^2 \,
   \Bigl[
    \grcp^2  \,
    F_0 ( 8\ttperp; q_0 \alpha )
\nonumber \\ && \hspace{-.1cm} +  \bigl( \grcm^2 + \grsp^2 +
\grsm^2 \bigr)  J_0 (q_0 \alpha )
  \Bigr] ,
\label{eq:Krho}
\\
\frac{d}{dl} K_{\sigma} &=& - \frac{1}{2} \, K_{\sigma}^2 \Bigl[
    \gpcp^2  \, J_0 (8\ttperp )
  + \gpcm^2
\nonumber \\ && \hspace{1.cm}
  + \gpsp^2
  + \gpsm^2
\Bigr] \virg
\\
\frac{d}{dl} \Kc &=& -\frac{1}{2} \sum_{p=\pm}
  \left[\Kc^2 \, J_0(8\ttperp) \, \delta_{p,+} - \delta_{p,-} \right]
\bigl(
  \gcsp^2
\nonumber \\ && \hspace{1.cm} + \gcsm^2 + \gpc^2 \bigr) \nonumber
\\ && -\frac{1}{2} \sum_{p=\pm}
  \bigl[\Kc^2 \, F_0(8\ttperp;q_0 \alpha) \, \delta_{p,+}
 \nonumber \\ && \hspace{1.cm}
         - J_0(q_0 \alpha) \, \delta_{p,-} \bigr]  \grc^2
\virg \label{eq:Kc}
\\
\frac{d}{dl}
 \Ks
&=& - \frac{1}{2} \sum_{p=\pm}
  \left(\Ks^2 \, \delta_{p,+} - \delta_{p,-} \right)
\Bigl[
    \grs^2 \, J_0 (q_0 \alpha )
\nonumber \\ && \hspace{.1cm}
  + \gps^2
  + \gcps^2 \, J_0 (8\ttperp ) + \gcms^2
\Bigr],
\end{eqnarray}%
\begin{eqnarray}
\frac{d}{dl} \gncpm &=& \left(2- \Knu - \Kc^{\pm} \right) \gncpm
\nonumber \\ && - \gnsp  \gcpmsp
- \gnsm  \gcpmsm ,
\\
\frac{d}{dl} \gnspm &=& \left(2- \Knu - \Ks^{\pm} \right) \gnspm
\nonumber \\ && - \gncp \, \gcpspm \, J_0(8\ttperp) \nonumber \\
&& - \gncm \, \gcmspm
\virg
\\
\frac{d}{dl} \gcs &=& \left(2- \Kc^{p} - \Ks^{p'} \right) \gcs
\nonumber \\ && - \grc \grsd \, J_0(q_0\alpha) \nonumber \\ && -
\gpc \gpsd \virg \label{eq:dl-gcs}
\end{eqnarray}
where $\nu=\rho,\sigma$ and   $p=\pm$. The quantities $F_0 (x;y)$
and $F_1 (x;y)$ are defined by
\begin{eqnarray*}
F_0 (x;y) &\equiv& \frac{1}{2}
   [J_0(|x+y|) + J_0(|x-y|)]
\virg
\\
F_1 (x;y) &\equiv& \frac{1}{2}
   [J_1(|x+y|) \, {\rm sgn}(x+y)
\nonumber \\ && + J_1(|x-y|) \, {\rm sgn}(x-y)] \virg
\end{eqnarray*}
 and $J_n$ is the $n$-th order Bessel function.
In the above equations,
 the $l$-dependence is not written explicitly.
 The  initial conditions  are given by
$K_\nu (0) = K_\nu$, $G_{\nu p, \nu' p'} (0) = g_{\nu p, \nu'
p'}/2\pi\vf$, $\ttperp (0) = \tperp/W$ with $W \equiv \vf
\alpha^{-1}$, and
 $q_0(0)= q_0$, respectively .
 The cutoff $\alpha$ can be related to the lattice constant
  by $\alpha=2a/\pi$ \cite{Tsuchiizu_PHB}, which leads to
 $W \simeq \pi t /\sqrt{2}$ for small $\td/t$.
 We take  $a$=1.
 The renormalization group equations for
  the velocity $v_\nu$  are discarded and
 the velocity $v_\rho$ and $v_\sigma$ are set to $\vf$.

In the limit of $t_{\perp} \rightarrow 0$ these equations reduce
to those of a single chain
 given by
\begin{eqnarray}
\frac{d}{dl} q_0\alpha
  &=& q_0 \alpha   -4 \, G_u^2 \, J_1(q_0\alpha)
\virg
\label{eq:rg1d-b}
\\
\frac{d}{dl} G_\rho &=& 2 \, G_u^2 \, J_0 (q_0 \alpha) \virg
\\
\frac{d}{dl} G_u &=& 2 \, G_\rho \, G_u \virg
\\
\frac{d}{dl} G_\sigma &=& 2 \, G_\perp^2 \virg
\\
\frac{d}{dl} G_\perp &=& 2 \, G_\sigma \, G_\perp \virg
\label{eq:rg1d-e}
\end{eqnarray}
  where $K_\rho = 1-G_\rho$ and $K_\sigma = 1-G_\sigma$.
In the above equations,
  $G_u \equiv G_{\rho+,C\pm}=\mp G_{\rho+,S\pm}$,
  $G_\perp \equiv \pm G_{\sigma+,C\pm}=-G_{\sigma+,S\pm}$,
  $(G_\rho+G_\sigma) \equiv \pm 2 G_{C\pm,S\pm}$ and
  $(G_\rho-G_\sigma) \equiv \mp 2 G_{C\pm,S\mp}$.
It is straightforward to derive
  Eqs.~(\ref{eq:rg1d-b})-(\ref{eq:rg1d-e})
  from
 the one-dimensional (1D) Hamiltonian  given by (Appendix A)
\begin{eqnarray}
{\cal H}_{\rm 1D} &=&
  \frac{v_\rho}{4\pi} \int dx
   \left[
      \frac{1}{K_\rho} \left(\partial_x \theta_+ \right)^2
    + K_\rho \left(\partial_x \theta_- \right)^2
   \right]
\nonumber \\ && + \frac{v_\sigma}{4\pi} \int dx
   \left[
      \frac{1}{K_\sigma} \left(\partial_x \phi_+ \right)^2
    + K_\sigma \left(\partial_x \phi_- \right)^2
   \right]
\nonumber \\ && -\frac{\mu}{\pi} \int dx \, \partial_x \theta_+
\nonumber \\ && + \frac{g_u}{2\pi^2 \alpha^2}  \int dx \,
  \cos 2\theta_+
\nonumber \\ && + \frac{g_\perp}{2\pi^2 \alpha^2}  \int dx \,
  \cos 2\phi_+
\virg
\label{eq:H1d}
\end{eqnarray}
where $\theta_\pm$ and $\phi_\pm$ are the phase variables
  expressing the charge and spin fluctuations, respectively
  \cite{Suzumura_PTP}.

\section{Commensurate and incommensurate states}

We examine the commensurate-incommensurate transition by solving the renormalization
group equations numerically. The dimerization  is taken as $\td/t=0.05$
where  such a choice does not change qualitatively the results as seen later.
The scaling quantity $l$ is  related to energy $\omega$ and/or temperature
by $l=\ln(W/\omega)=\ln(W/T)$.

\begin{figure}[t]
\epsfxsize=3.5in\epsfbox{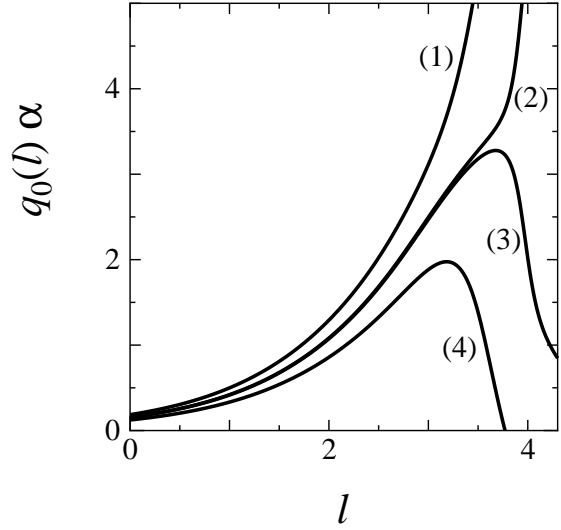} \caption{
 The scaling flows of $q_0(l)\alpha$
  with fixed $\mu/t=0.15$(1), 0.126(2), 0.125(3) and 0.1(4) for
  $U/t=5$, $\td/t=0.05$ and $t_\perp/t=0.1$.
}
\end{figure}
 Equation~(\ref{eq:dl-q0}) shows that
   the quantity $q_0(l)$ in the absence of the interaction
      increases as $q_0(l)=q_0\e^l$,
 while the increase of $q_0(l)$ is suppressed by
   the presence of   $G_{\rho +, C\pm}$ and $G_{\rho +, S\pm}$
   coming from the umklapp scattering $g_u$.
In Fig.~1,     $q_0(l)$ are shown with
 several choices of $\mu/t$
  for   $U/t=5$, $\td/t=0.05$ and $\tperp/t=0.1$.
For $\mu/t=0.15$ (curve (1)),
  the misfit term $q_0(l)$ increases rapidly
  and becomes relevant. In this case,
  the umklapp scattering can be neglected
  since   the Bessel functions
  $J_0(q_0\alpha)$ and $J_1(q_0\alpha)$
   in Eqs.~(\ref{eq:dl-tperp})-(\ref{eq:dl-gcs})
   become  small due to  large $q_0\alpha$.
The quantity $K_\rho(l)$  at the fixed point   takes a finite
value due to the relevant $q_0 \alpha$,
  and then  the incommensurate state
is obtained. For smaller value of the chemical potential given by
$\mu/t=0.1$ (curve(4)),
 the quantity $q_0(l)\alpha$ has a maximum
  and reduces to zero with increasing $l$,
        showing the irrelevant  $q_0(l)\alpha$.
Such a behavior of $q_0(l)\alpha$ leads to the commensurate state,
 which gives an insulating state due to
  the zero value of   $K_\rho(l)$
     in  the limit  of large $l$.
We note that the commensurate state
   corresponds to that of  a half-filled case.
 The  commensurate-incommensurate transition occurs
  at the critical value given by $\mu/t=\mu_c/t (\simeq 0.125)$
  (curve (3)).

\subsection{Half-filled case}

\begin{figure}[t]
\epsfxsize=3.5in\epsfbox{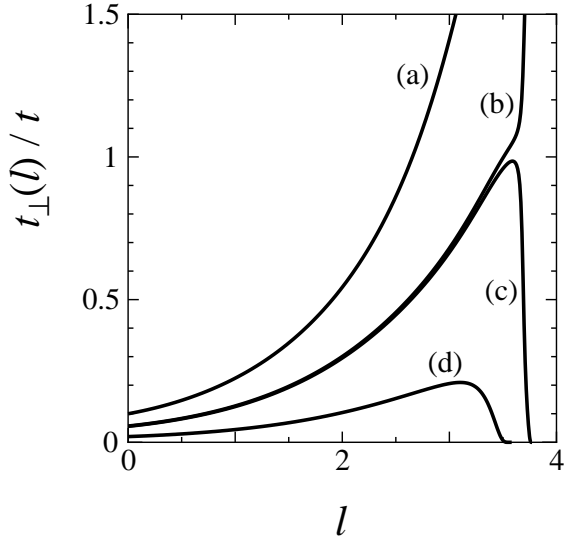} \caption{
The scaling flows of
$\tperp(l)/t$
  with fixed $\tperp/t=0.1(a)$, 0.057(b), 0.056(c) and 0.02(d)
  for $\mu/t=0.1$,
 $U/t=5$ and  $\td/t=0.05$.
}
\end{figure}
It is known that a transition from the relevant interchain hopping
to the irrelevant one occurs with increasing $t_{\perp}$
 at half-filling \cite{Suzumura,Tsuchiizu}.
 We show that such a result is also obtained for $\mu \not= 0$,
which still leads to the commensurate state.
 In  Fig.~2, the quantity $\tperp(l)/t$ is
  calculated  for the  commensurate state with  $\mu/t=0.1$.
    With decreasing the interchain hopping as
     $\tperp/t=0.1$(a), 0.057(b), 0.056(c) and 0.02(d),
   the hopping changes from a divergent  behavior
    to a convergent one with  zero value.
 One finds the relevant interchain hopping for
 large $t_{\perp}$ (curve (a))
  but irrelevant one for small $t_{\perp}$ (curve (d)).
 The behavior which separates these two cases is obtained
 at a critical value of $t_{\perp} = t_{\perp, c}$ (curve(c)).
For a relevant interchain hopping,
   one obtains an insulating phase with spin gap excitations
which is  called  ``D-Mott'' phase \cite{lin_so8} with
``C0S0'' phase. The notation C$n$S$m$
denotes a state with $n$ massless charge mode and $m$ massless
spin mode.
Such a spin gap is also known in the $SO(8)$ Gross-Neveu model
   \cite{lin_so8}, which
 includes an additional term given by
   $g_{\rho+,\sigma+} \, \sin \sqrt{2}\theta_{\rho+} \,
                         \cos \sqrt{2}\theta_{\sigma+}$.
For an irrelevant interchain hopping,
   one can also expect a state  with the spin gap
   due to the following fact.
The phase has all charge excitations gapped,
   and because of such generated Heisenberg exchange
   ($J\simeq \tperp^2/\Delta_\rho$) \cite{LeHur}
   is equivalent to a spin ladder.
Such a system is known to have all spin excitations gapped.
We note that, for two-coupled chains, these states
 undergo a crossover since
   both states exhibit C0S0 phase and
   there is no clear distinction between  these two states
   \cite{LeHur,Donohue}.
However we may identify the irrelevant (relevant) $t_{\perp}$
   as  confinement  (deconfinement) since such a result could
   be expected    for the case of many chains
  \cite{Suzumura_JPCS}  and higher dimension.

\subsection{Commensurate-incommensurate transition}

\begin{figure}[t]
\epsfxsize=3.5in\epsfbox{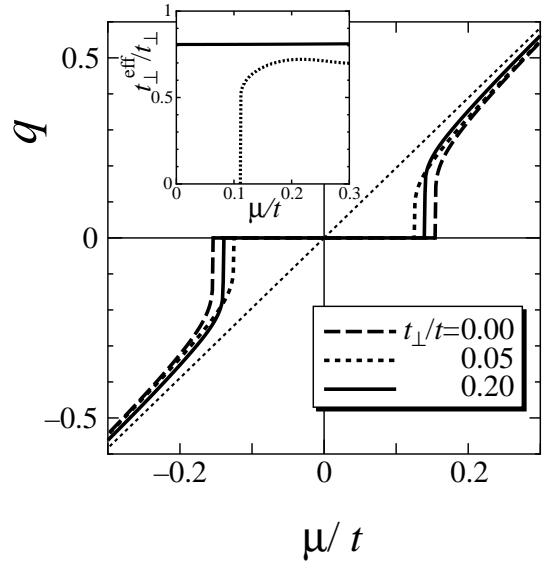} \caption{
The effective
quantity $q$ vs. the chemical potential $\mu/t$
  with fixed $\tperp/t=0$, 0.05 and 0.2 for $U/t=5$ and
  $\td/t=0.05$.
The thin dotted line denotes $q_0$. The inset shows
 $\tperp^{\rm eff}/\tperp$,
  for $\tperp/t=0.05$(dotted curve) and 0.20 (dashed curve).
}
\end{figure}
 Based on  Figs.~1 and 2, we examine
  quantities $q$ and $t_{\perp}^{\rm eff}$, which express
       effective quantities  of $q_0$ and $\tperp$
    in the presence of the interaction and the chemical potential.
  These quantities  are   estimated from $q\alpha=\exp[-l_{q}]\times 4$
      with $q_0(l_q)\alpha=4$,
  and $t_\perp^{\rm eff}=t\exp[-l_1]$ with $\tperp(l_1)/t=1$.
 One finds that  $q=q_0$ and $\tperp^{\rm eff}=\tperp$ for
     the noninteracting case.
In the present analysis,
 the effective quantity $q$ is  related to the
  carrier density $n$ by $n=q/\pi$ (Appendix A).
In Fig.~3, the  $\mu$-dependence of $q$
  is shown with fixed $\tperp/t=0$, 0.05 and 0.2 for
   $U/t=5$ and $\td/t=0.05$.
The quantity $q$, which is  suppressed by the umklapp scattering,
   becomes zero in the commensurate phase.
 The thin dotted line denotes  $q (= q_0)$
  in the absence of the umklapp scattering.
  The critical value $\mu_c$ for the commensurate-incommensurate
  transition is not monotonical as a function of $t_{\perp}$ since
    $\mu_c$  for  $\tperp/t=0 $ is larger (smaller)
    than that for  $\tperp/t=0.05$ (0.2).
 The explicit $\tperp$-dependence of $\mu_{c}$
    is evaluated in Fig.~4.
In the inset of Fig.~3, the effective interchain hopping
normalized
 by $t_{\perp}$
    is shown for $\tperp/t=0.05$ (dotted curve) and 0.2 (solid curve)
 where  $\tperp^{\rm eff}/\tperp$ is
     symmetric with respect to $\mu=0$.
For $\tperp/t=0.05$,
    $\tperp^{\rm eff}/\tperp$ becomes zero for small $\mu/t (\lsim  0.11)$
  and exhibits  the irrelevant interchain hopping
  while the interchain hopping for $\tperp/t=0.2$
  shows always  relevant one.

\begin{figure}[t]
\epsfxsize=3.5in\epsfbox{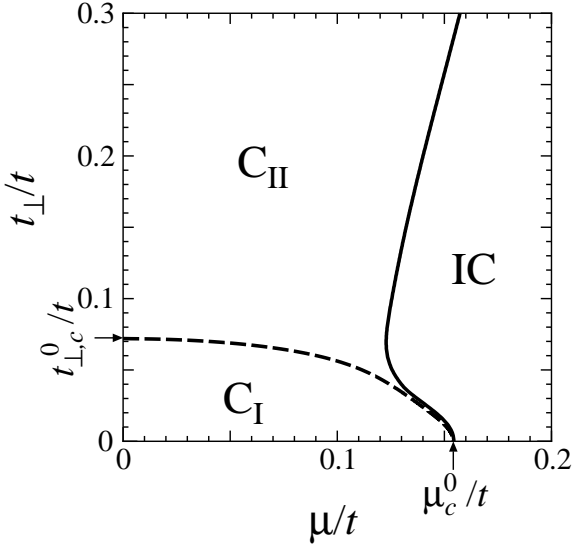} \caption{
The phase diagram of
commensurate (C$_{\rm I}$ and C$_{\rm II}$) and  incommensurate
(IC) states
  on the plane of $\mu/t$ and $\tperp/t$
    for $U/t=5$ and $\td/t=0.05$.
The region
 C$_{\rm I}$ (C$_{\rm II}$) denotes
  ``Confined phase'' (``D-Mott phase''), i.e.,
 the commensurate state with the irrelevant
(relevant) interchain hopping.
The region IC corresponds to ``Luther-Emery liquid''.
The arrow on the horizontal (vertical) axis shows
 the critical value $\mu_{c}^0/t$ ($t_{\perp,c}^0/t$)
  separating the respective regions.
}
\end{figure}
From the results of Fig.~3, we examine
  the boundary between
  the  commensurate state (C$_{\rm I}$ and C$_{\rm II}$)
    and the incommensurate state (IC)
   and also the boundary between the
     relevant interchain hopping and irrelevant one.
 In Fig.~4, the phase diagram of these states
 is shown for  $U/t=5$ and $\td/t=0.05$.
The solid curve denotes the boundary between
 the incommensurate state and the commensurate state.
 The incommensurate state
 corresponds to the metallic state.
 The commensurate state is the (Mott-)insulating state
 where
 the dashed curve in the commensurate state denotes
 the boundary between the  relevant interchain hopping
 (``D-Mott phase'') \cite{LeHur}
  and the irrelevant one (``Confined phase'').
 The  critical value shown by the arrow on the  $t_{\perp}$-axis
   is given by   $t_{\perp,c}^0/t \simeq 0.072$
\cite{Tsuchiizu_PHB,Suzumura_JPCS}.
 The critical value shown by the arrow on the $\mu$-axis
 is given by  $\mu_{c}^0/t \simeq 0.154$, at which
 the solid curve merges with the dashed curve.
 The interchain hopping is always relevant in the incommensurate phase
      since the umklapp scattering
   becomes irrelevant  in the limit of  low energy
    as seen from   Eq.~(\ref{phase_Hamiltonian}).
The incommensurate phase shows  no  gap in the total charge
fluctuation
due to the relevant $\mu$ \cite{KR},
 while the gaps
still exist for other fluctuations
due to the relevant $\tperp$.
The incommensurate state  corresponds to a  ``C1S0''
$d$-wave superconductivity \cite{Schulz} and is called
   ``Luther-Emery liquid'' \cite{Ledermann}.
The solid curve corresponding to the critical value
 $\mu_c$  shows non-monotonical behavior
 as a function of  $\tperp$.
With increasing $t_{\perp}$,
   $\mu_c$ has a minimum   at $\tperp/t\simeq 0.07$
     and becomes larger than $\mu_c^0$.
 The decrease of $\mu_c$ for  small $\tperp/t(\lsim 0.07)$
 originates in the fact that
     the effect of  the umklapp scattering
       becomes weakened  for finite $\tperp$ due to the misfit
   $(8\tperp/\vf)x$ in  $g_u$-term of  Eq.~(\ref{phase_Hamiltonian}).

For large $t_\perp$ and small $g_u$ ($g_u$ is controlled by $\td$
which is small here) we can explain the behavior by a qualitative
analysis of the renormalization group equations.
The strong interchain hopping
between the two interacting chains opens a gap in all sectors
except for the total charge sector.
We may define a scale $l_1$ where
the couplings in the $\sigma$, $C$, $S$ sectors have reached a
value of order one. This scale will depend on $t_\perp$ and the
value of the bare interaction but not on $g_u$. Up to $l_1$, $g_u$
will renormalize by some finite multiplicative constant that will
not affect the asymptotic dependence of the charge gap. However
above $l_1$ the scaling dimension of $g_u$ is $2-K_\rho$, instead
of $2-2K_\rho$ if $t_\perp=0$. Indeed $\cos(\sqrt2\theta_{S+})$
and $\cos(\sqrt2\theta_{C-})$ acquire a mean value (with opposite
signs) so that the umklapp term reduces to:
\begin{equation} \label{eq:umklappeff}
g_u\sin\sqrt2\theta_{\rho+}
  \left(\langle\cos\sqrt2\theta_{C-}\rangle
-\langle\cos\sqrt2\theta_{S+}\rangle \right)
\point
\end{equation}
As a first consequence this means that the power law dependence of
the gap for asymptotically small umklapp scattering is enhanced by
interchain hopping from $\Delta_\rho^{\rm 1D}\propto
g_u^{1/(2-2K_\rho)}$ to $\Delta_\rho\propto g_u^{1/(2-K_\rho)}$.
The second physical consequence is for the
commensurate-incommensurate transition.
In the absence of chemical potential of Eq.~(\ref{eq:mu}),
the operator $g_u$ in Eq.~(\ref{eq:umklappeff}) would be relevant when
$K_{\rho}^{\mu=0}<2$. As is standard for the commensurate-incommensurate
transition \cite{Schulz,Giamarchi_physica},
the addition of a chemical potential of Eq.~(\ref{eq:mu})
leads to a universal value of the Luttinger exponent close to zero
doping of $K^*_{\rho}=K_{\rho}^{\mu=0}/2=1$. It is also easy to
see that the charge excitations connecting two minima of the potential,
Eq.~(\ref{eq:umklappeff}), correspond to a charge $+2$.
We thus recover quite generally the results that were established
in the various previous limits \cite{konik_exact_commensurate_ladder,schulz_mitwochain,%
Ledermann,Siller}.

\begin{figure}[t]
\epsfxsize=3.5in\epsfbox{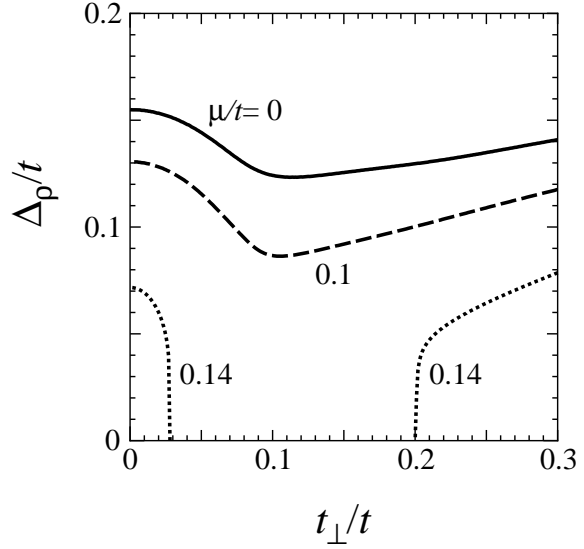} \caption{
The
$t_\perp$-dependence of the charge gap $\Delta_\rho/t$
for fixed  $\mu /t= 0$, 0.1 and 0.14 for $U/t=5$ and $\td/t=0.05$.
}
\end{figure}
 Next,  we estimate  the magnitude of the charge gap on the plane of
  $\mu/t$ and $t_{\perp}/t$ of Fig.~4.
In a way similar to the previous calculations in  the case
 for  $\mu=0$
\cite{Suzumura,Tsuchiizu},
 the charge gap $\Delta_{\rho}$   is calculated from the
 renormalization group flow
 where     $\Delta_\rho/t=W \exp[-l_\Delta]$
   and  $l_\Delta$  is defined    by  $K_\rho (l_\Delta) \equiv 1/2$.
 We  set $\Delta_\rho=0$ for $K_\rho (\infty) > 1/2$.
 In Fig.~5, the $\tperp$-dependence of $\Delta_{\rho}$
 is  shown  for $\mu/t=0$, 0.1 and 0.14.
 When $t_{\perp}$ increases,
    the charge gap with $\mu = 0$   decreases  and has a minimum at
  $t_{\perp}/t = 0.154$.
 A similar dependence is found in the $t_{\perp}$-dependence of
   $\mu_c$ in Fig.~4 where the location for the minimum is the same
  within the numerical accuracy. The identification of the
charge gap for the commensurate case $\Delta_\rho(\mu=0)$ with $\mu_c$ can
be expected for the commensurate-incommensurate transition on general
grounds. Indeed the charge gap is the
smallest energy one
must pay to inject an excitation which carries
a charge. When the chemical potential reaches the charge
gap, charged excitations are injected in the system, the system is
incommensurate.
The non-monotonical dependences of $\Delta_\rho$ on $t_\perp$
indicate a crossover
    from the irrelevant interchain
 hopping to the relevant one with increasing $t_{\perp}$.
 For  $\mu/t=0.1$, $\Delta_{\rho}$ is suppressed but is
  still   similar  to that of  $\mu=0$.
However the $t_{\perp}$-dependence of $\Delta_{\rho}$
 for $\mu/t=0.14$  is  qualitatively  different from others.
 The   metallic state with  $\Delta_\rho/t=0$
 appears   in the interval region
   of $0.028 < \tperp/t < 0.2$.
 This region  corresponds to
     the incommensurate state  in  Fig.~4.
We note that numerical integration of the
renormalization group flow shows
that  $\Delta_{\rho}$ as a function of $\mu$
decreases monotonically to zero
at the commensurate-incommensurate transition, in agreement
with the annihilation of the gap
expected for a commensurate-incommensurate transition.

\begin{figure}[t]
\epsfxsize=3.5in\epsfbox{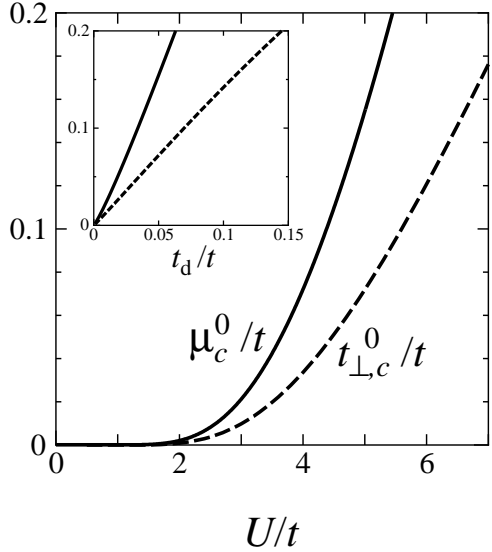} \caption{
 The $U/t$-dependence of $\mu_c^0$ (solid curve) and
  $t_{\perp,c}^0$ (the dashed curve) for $\td/t=0.05$,
    which are shown by the arrows of the horizontal axis
 and the vertical axis in Fig.~4, respectively.
 The inset shows the corresponding $\td$-dependence
  for $U/t=5$.
}
\end{figure}
Finally  we examine the $U$ and $\td$
   dependences  of the critical values,
   $\mu_c^0$ and $t_{\perp,c}^0$,  which are
       shown by the arrows in Fig.~4.
 In Fig.~6
 the $U$-dependences of  $\mu_c^0$   and $t_{\perp,c}$
   are shown by the solid curve and the dashed curve respectively
  for  $\td/t=0.05$.
 In the inset, the $\td$-dependences
   of the corresponding quantities are shown
    by  solid and  dashed curves    for $U/t=5$.
 These $U$ and $\td$-dependences show
   that   $\mu_c^0 / t_{\perp,c}^0 \simeq 2.2$
     for the present choice of parameters.
Thus a rescaled phase diagram, which is
independent of $U$ and $\td$, can be
obtained  from  Fig.~4 by using the rescaled variables
$\mu/\mu_{c}^0$ and $t_{\perp}/t_{\perp,c}^0$. Our renormalization
group procedure correctly gives a value for $\mu_c^0$ identical,
within the scale of Fig.~4, with $\Delta_\rho^{\rm 1D}$.
The quantity $\Delta_\rho^{\rm 1D}$ denotes
$\Delta_{\rho}$ of the single chain where
  $\Delta_\rho^{\rm 1D} \simeq W(g_u/W)^{1/(2-2K_\rho)}$
 for small $g_u$
\cite{Schwartz}.
 We note  that the  boundary between a relevant $t_{\perp}$
 and an irrelevant $t_{\perp}$ is determined more accurately
  by the competition between
   $\tperp^{\rm eff,0}$ and  $\Delta_\rho^{\rm 1D}$ where
  $\tperp^{\rm eff,0}$ denotes the effective   interchain hopping energy
  for $\td = \mu = 0$
  \cite{Tsuchiizu_PRG,Suzumura_JPCS}.
  The quantity $\tperp^{\rm eff,0}$ is given analytically by
  $\tperp^{\rm eff,0}= \tperp (\tperp/t)^{\alpha_0/(1-\alpha_0)}$
   with
  $\alpha_0=(K_\rho+K_\rho^{-1}+K_\sigma+K_\sigma^{-1}-4)/4$
  \cite{Bourbonnais1,Bourbonnais2,Bourbonnais}.

\section{Summary and discussion}

In the present paper, we investigated
the properties of two-coupled chains  with
interchain electron hopping and a filling  close to half-filling.
By using bosonization and a renormalization group method we obtained the full phase
diagram of the system. There is a metal-insulator transition
for a critical value $\mu_c$ of the chemical potential, describable by
a commensurate-incommensurate phase transition on the bosonized Hamiltonian.
The critical value $\mu_{c}$, shown in Fig.~4, exhibits a non-monotonical  dependence
on  the perpendicular hopping $\tperp$. The minimum of $\mu_c$ occurs for
values of $t_\perp$ close to the ones for which a confinement deconfinement
crossover takes place for the commensurate case. For large $t_\perp$
the relevance of interchain hopping reinforces the commensurate character of the
system leading to an enhanced commensurability gap.
The crossover line separating the regions of irrelevant and relevant
interchain hopping (confinement-deconfinement line)
merges with the boundary between the commensurate
state and the incommensurate state at $\mu=\mu_c^0$, the critical chemical
potential for a single chain ($t_\perp=0$).
We found that the phase diagram of Fig.~4 becomes almost independent of parameters
such as interactions and dimerization when expressed in terms of the
rescaled variables $\mu/\mu_c^0$ ($t_{\perp}/t_{\perp,c}^0$).
In addition, given the form for the Hamiltonian, we could show that
at the limit of small doping, the Luttinger liquid parameter takes
the universal value $K_\rho^*=1$, thereby confirming the
results
\cite{lin_so8,konik_exact_commensurate_ladder,schulz_mitwochain,%
      Ledermann,Siller}
obtained on specific limits of the model.

Finally let us discuss
the interchain exchange interactions in
the commensurate confined phase (irrelevant
interchain hopping).
Even when irrelevant, the interchain hopping
generates two particles and particle hole hopping \cite{Bourbonnais}.
Within the present formalism, the two particle interchain hopping
can be taken into account by starting from the  chain.
The Hamiltonian corresponding to the umklapp scattering,
${\cal H}_u$, in Eq.~(\ref{phase_Hamiltonian}) is rewritten as
\begin{eqnarray}
{\cal H}_u
 &=&
g_u^{\rm 1D} \sum_{p,l} \int dx \,
  \psi_{p,\uparrow,l}^\dagger   \, \psi_{-p,\uparrow,l}  \,
  \psi_{p,\downarrow,l}^\dagger \, \psi_{-p,\downarrow,l}
\nonumber \\ &&+ J_{uz} \sum_p \int \! dx \left[
    \psi_{p,\uparrow,1}^\dagger   \, \psi_{-p,\uparrow,1} \,
    \psi_{p,\downarrow,2}^\dagger \, \psi_{-p,\downarrow,2}
  + {\rm h.c.}
\right] \nonumber \\ &&+ J_{u\perp} \sum_p \int  \! dx \left[
    \psi_{p,\uparrow,1}^\dagger   \, \psi_{-p,\downarrow,1} \,
    \psi_{p,\downarrow,2}^\dagger \, \psi_{-p,\uparrow,2}
  + {\rm h.c.}
\right] \nonumber \\ && + g_u^{pt} \sum_p  \int  \! dx \left[
    \psi_{p,\uparrow,1}^\dagger   \, \psi_{p,\downarrow,1}^\dagger \,
    \psi_{-p,\downarrow,2} \, \psi_{-p,\uparrow,2}
  + {\rm h.c.}
\right], \nonumber \\
\end{eqnarray}
  where $\psi_{p,\sigma,l}=L^{-1/2} \sum_k \e^{i k x} d_{k,p,\sigma,l}$
  with $l=1,2$ denoting the chain index and $p=+(-)$ corresponding to
  the right (left) moving electrons.
The coupling constants are given by
 $g_u^{\rm 1D}=(g_{\rho+,C+}+g_{\rho+,C-}-g_{\rho+,S+}+g_{\rho+,S-})/4$,
 $J_{uz}=(g_{\rho+,C+}-g_{\rho+,C-}-g_{\rho+,S+}-g_{\rho+,S-})/4$,
 $J_{u\perp}=-(g_{\rho+,C+}-g_{\rho+,C-}+g_{\rho+,S+}+g_{\rho+,S-})/4$
  and
 $g_u^{pt}=(g_{\rho+,C+}+g_{\rho+,C-}+g_{\rho+,S+}-g_{\rho+,S-})/4$,
  where $g_u^{\rm 1D}$ and $J_{uz}$ ($J_{u\perp}$) denote
  the intrachain  umklapp scattering and
  the interchain umklapp exchange, and $g_u^{pt}$ denotes the pair
  tunneling between chains.
Other terms in the Hamiltonian can be rewritten in a similar
form. The initial values of the interchain exchange  and that of
pair tunneling are zero since the initial values
of the umklapp scattering in the renormalization group equations
are given by $g_{\rho +,C+}=g_{\rho+,C-}=-g_{\rho+,S+}=g_{\rho+,S-}=g_u$.
Both of these interactions are  generated by the renormalization.

It would be a interesting problem to investigate, by taking the higher
order terms in the renormalization group, the consequences
of these terms on the full phase diagram investigated in this paper.

\begin{acknowledgement}
One of the authors (Y.S.) is thankful for the financial support
 from Universit{\'e} Paris--Sud and also for the kind hospitality
   during his stay at Ecole Normale  Sup\'erieure.
M.T.  and Y.S.  thank T. Itakura for useful discussions. This work
was partially supported by a Grant-in-Aid for Scientific
  Research from the Ministry of Education, Science, Sports and
  Culture (Grant No.09640429), Japan.
\end{acknowledgement}

\appendix

\section{Renormalization group equations }

In this section, we derive
the renormalization group equation for $\mu$.

First, we treat the system with
a single chain where the phase Hamiltonian is given by
\begin{eqnarray}
{\cal H}_{\rm 1D} &=&
  \frac{v_\rho}{4\pi} \int dx
   \left[
      \frac{1}{K_\rho} \left(\partial_x \theta_+ \right)^2
    + K_\rho \left(\partial_x \theta_- \right)^2
   \right]
\nonumber \\ &&
+ \frac{g_u}{2\pi^2 \alpha^2}  \int dx \,
  \cos \left(2\theta_+ + q_0 x \right)
\point
\end{eqnarray}
The quantities $v_\rho$ and $K_\rho$ are the same as two-coupled
chains,
  and $[\theta_+(x),\theta_-(x')]=i\pi {\rm sgn}(x-x')$
  \cite{Suzumura_PTP} and
  $q_0=(4K_\rho/v_\rho)\mu$.
Here the expectation value of  the carrier density, $n$,  can be
evaluated  as
\begin{eqnarray}
n &=& \frac{2K_\rho}{\pi v_\rho} \mu
  + \frac{1}{\pi}\frac{T}{L} \int dx \, d\tau \,
     \lan  \partial_x \theta_+ \ran
\virg \label{eq:n}
\end{eqnarray}
where the factor $2K_\rho/\pi v_\rho$ in the first term of r.h.s.
  corresponds to the compressibility in the absence of
  the umklapp scattering.
The second term of Eq.~(\ref{eq:n}) can be calculated as follows,
\begin{eqnarray}
&& \int dx \, d\tau \,
     \lan  \partial_x \theta_+ \ran
\nonumber \\ &=& \frac{1}{Z} {\rm Tr}
  \left[\left.
    \frac{\partial}{\partial \lambda}
      \exp \left(
              - \int d\tau \, {\cal H}_{\rm 1D}
                       + \lambda \int dx \, d\tau
                  \, \partial_x \theta_+
           \right)  \right|_{\lambda=0}
  \right]
\nonumber \\ &=& \frac{4G_u}{\alpha^2} K_\rho
  \int dx \, d\tau \, x \lan \sin\left(2\theta_+ + q_0 x \right)  \ran
\virg \label{eq:ex-theta}
\end{eqnarray}
where $Z={\rm Tr} \, \exp(-\int d\tau \, {\cal H}_{\rm 1D})$
   and $G_u=g_u/(2\pi v_\rho)$.
In Eq.~(\ref{eq:ex-theta}),
  the new phase variable
  $\widetilde{\theta}_+(= \theta_+ - 2\pi K_\rho \lambda x/v_\rho)$
  has been introduced
  and  rewritten as $\widetilde{\theta}_+ \to \theta_+$.
Then Eq.~(\ref{eq:n}) leads
\begin{eqnarray}
n &=& \frac{1}{2\pi} q_0
  + \frac{4}{\pi \alpha^2}  G_u K_\rho \frac{T}{L} \int dx \, d\tau \,
      x \lan \sin\left(2\theta_+ + q_0 x \right)  \ran
. \nonumber \\ \label{eq:nb}
\end{eqnarray}
Here it is worthwhile noting that
  Eq.~(\ref{eq:nb}) is compared with
\begin{eqnarray}
\mu &=& \frac{\pi v_\rho}{2K_\rho} n
  - \frac{2 v_\rho}{\alpha^2}  G_u \frac{T}{L} \int dx \, d\tau \,
      x \lan \sin\left(2\theta_+ + 2\pi n x \right)  \ran
,
\nonumber \\
\label{eq:ap-mu}
\end{eqnarray}
  which is obtained
  by using the Legendre transformation, i.e.,
  by calculating the value of the chemical potential at fixed
  carrier density $n$
  \cite{Giamarchi_JPF,Giamarchi_PRB}.

After a straightforward calculation of Eq.~(\ref{eq:nb}), one
finds, up to the lowest order of $G_u$,
\begin{eqnarray}
n  &=& \frac{1}{2\pi} q_0
  - \frac{2}{\pi \alpha}  G_u^2 K_\rho \int
   \frac{dr}{\alpha} \left(\frac{r}{\alpha}\right)^{2-4K_\rho}
  \! J_1(q_0 r).
\label{eq:ap-n}
\end{eqnarray}
By assuming that the quantity $n$  is simply scaled as
  $n(l)=n\, \e^l$ \cite{Giamarchi_JPF} with the transform
  $\alpha\to \alpha\, \e^{l}$,
  the renormalization
  group equation for $q_0$ is obtained as,
\begin{eqnarray}
\frac{d}{dl} q_0 \alpha
 &=&  q_0 \alpha
 -
4 G_u^2 \, K_\rho \, J_1 (q_0 \alpha) \point
\end{eqnarray}
The renormalization group equations for $K_\rho$ and $G_u$
  can be obtained in a way similar to
  Ref.~\cite{Giamarchi_PRB} as
\begin{eqnarray}
\frac{d}{dl} K_\rho
 &=&  - 2 \, G_u^2 \, K_\rho^2 \, J_0(q_0 \alpha)
\virg \\ \frac{d}{dl}G_u &=& (2-2K_\rho) \, G_u \point
\end{eqnarray}
By integrating this renormalization group equation, the
  effective quantity of $q_0$ can be estimated from
   $q\alpha=c \exp(-l_{q})$
   with $q_0(l_q)\alpha=c$, where $c$ is numerical constant of
  the order of unity.
One finds that the  quantity $q$ can be related to
  carrier density $n$ by $n=q/(2\pi)$ from Eq.~(\ref{eq:ap-n}).

Next we calculate
  the case of two-coupled chains.
The expectation value of the carrier density, $n$, can be evaluated  as
\begin{eqnarray}
n &=& \frac{4K_\rho}{\pi v_\rho} \mu
  + \frac{\sqrt{2}}{\pi}\frac{T}{L} \int dx \, d\tau \,
     \lan  \partial_x \theta_{\rho +} \ran
\point \label{eq:n2}
\end{eqnarray}
  where the first term of r.h.s.~becomes twice
  as that in Eq.~(\ref{eq:n}),
  since here we consider two chains.
 From a procedure similar to the single chain,
  Eq.~(\ref{eq:nb}) is replaced by
\begin{eqnarray}
n&=& \frac{1}{\pi} q_0 + \frac{4}{\pi \alpha^2} G_{\rho+,C+}\,
K_\rho
  \frac{T}{L}\int dx\, d\tau \, x 
\nonumber \\ &&  \hspace{.5cm} \times
\lan
   \sin \left(\sqrt{2}\theta_{\rho +}+q_0 x\right) \,
   \cos \left(\sqrt{2}\theta_{C +}-\frac{8\tperp}{\vf}x\right)
\ran \nonumber \\
&+&  \frac{4}{\pi \alpha^2} G_{\rho+,C-}\,
K_\rho
  \frac{T}{L}\int dx\, d\tau \, x
\nonumber \\ &&  \hspace{1.cm} \times
\lan
   \sin \left(\sqrt{2}\theta_{\rho +}+q_0 x\right)  \,
   \cos \sqrt{2}\theta_{C -}
\ran \nonumber \\ &+&  \frac{4}{\pi \alpha^2} G_{\rho+,S+}\,
K_\rho
  \frac{T}{L}\int dx\, d\tau \, x
\nonumber \\ &&  \hspace{1.cm} \times
\lan
   \sin \left(\sqrt{2}\theta_{\rho +}+q_0 x\right)  \,
   \cos \sqrt{2}\theta_{S +}
\ran \nonumber \\ &+&  \frac{4}{\pi \alpha^2} G_{\rho+,S-}\,
K_\rho
  \frac{T}{L}\int dx\, d\tau \, x
\nonumber \\ &&  \hspace{1.cm} \times
\lan
   \sin \left(\sqrt{2}\theta_{\rho +}+q_0 x\right)  \,
   \cos \sqrt{2}\theta_{S -}
\ran \point \label{eq:n2b}
\end{eqnarray}
The coupling constants that appear in Eq.~(\ref{eq:n2b})
  are those including the misfit $q_0 x$ in the cosine potential
  of Eq.~(\ref{phase_Hamiltonian}).
 The calculation in  the lowest order of  perturbation yields,
\begin{eqnarray}
n & = &
 \frac{1}{\pi} q_0
+ \frac{1}{\pi\alpha} G_{\rho+,C+}^2 \, K_\rho
\nonumber \\ &&  \hspace{1.cm} \times
  \int \frac{dr}{\alpha}
  \left(\frac{r}{\alpha}\right)^{2-2K_\rho -2 K_C} \,
  F_1\left(q_0 r; 8t_\perp r/\vf \right)
\nonumber \\ && + \frac{1}{\pi \alpha} G_{\rho+,C-}^2 \, K_\rho
  \int \frac{dr}{\alpha}
  \left(\frac{r}{\alpha}\right)^{2-2K_\rho -2/K_C} \,
  J_1(q_0 r)
\nonumber \\ && + \frac{1}{\pi \alpha} G_{\rho+,S+}^2 \, K_\rho
  \int \frac{dr}{\alpha}
  \left(\frac{r}{\alpha}\right)^{2-2K_\rho -2K_S} \,
  J_1(q_0 r)
\nonumber \\ && + \frac{1}{\pi \alpha} G_{\rho+,S-}^2 \, K_\rho
  \int \frac{dr}{\alpha}
  \left(\frac{r}{\alpha}\right)^{2-2K_\rho -2/K_S} \,
  J_1(q_0 r)
\virg
\nonumber \\
\label{eq:n2c}
\end{eqnarray}
where $F_1 (x;y) \equiv
   [J_1(|x+y|) \, {\rm sgn}(x+y)
    + J_1(|x-y|) \, {\rm sgn}(x-y)]/2$.
The infinitesimal transform of the cutoff
  $\alpha\to \alpha'=\alpha\,e^{dl}$ in Eq.~(\ref{eq:n2c})
  leads the renormalization equation, Eq.~(\ref{eq:dl-q0}).
  The renormalization group equation for $\tperp$,
  Eq.~(\ref{eq:dl-tperp}), can be obtained in a similar way.
The equations for the other
  coupling constants, Eqs.~(\ref{eq:Krho})-(\ref{eq:dl-gcs})
   are also obtained in a way similar to
  Ref.~\cite{Tsuchiizu}.

\end{document}